\documentclass[conference]{IEEEtran}

\usepackage{graphicx}
\usepackage{amsmath} 
\usepackage{xspace}
\usepackage{url}

\newcommand{\etal}{\emph{et al.}\xspace}
\newcommand{\ie}{i.e.,\xspace}
\newcommand{\eg}{e.g.,\xspace}
\newcommand{\fig}[1]{Figure~\ref{#1}}
\newcommand{\tab}[1]{Table~\ref{#1}}

\newcommand{\github}{GitHub\xspace}

\newcommand{\java}{Java\xspace}
\newcommand{\javascript}{JavaScript\xspace}

\newcommand{\npm}{npm\xspace}

\newcommand{\maven}{Maven\xspace}

\newcommand{\librariesio}{libraries.io\xspace}
\newcommand{\npmsio}{npms.io\xspace}
\newcommand{\npmjs}{npmjs.com\xspace}

\begin{document}

\title{On the Diversity of Software Package Popularity Metrics: An Empirical Study of npm}

	\author{\IEEEauthorblockN{Ahmed Zerouali}
	\IEEEauthorblockA{UMONS, Belgium}
	\IEEEauthorblockA{ahmed.zerouali@umons.ac.be}
	\and
	\IEEEauthorblockN{Tom Mens}
	\IEEEauthorblockA{UMONS, Belgium}
	\IEEEauthorblockA{tom.mens@umons.ac.be}
	\and
	\IEEEauthorblockN{Gregorio Robles }
	\IEEEauthorblockA{URJC, Spain}
	\IEEEauthorblockA{grex@gsyc.urjc.es}
	\and
	\IEEEauthorblockN{Jesus M. Gonzalez-Barahona}
	\IEEEauthorblockA{URJC, Spain}
	\IEEEauthorblockA{jgb@gsyc.es}	
}

\maketitle

\begin{abstract}
Software systems often leverage on open source software libraries to reuse functionalities. Such libraries are readily available through software package managers like \npm for \javascript. 
Due to the huge amount of packages available in such package distributions, developers often decide to rely on or contribute to a software package based on its popularity. Moreover, it is a common practice for researchers to depend on popularity metrics for data sampling and choosing the right candidates for their studies. However, the meaning of popularity is relative and can be defined and measured in a diversity of ways, that might produce different outcomes even when considered for the same studies. In this paper, we show evidence of how different is the meaning of popularity in software engineering research.
Moreover, we empirically analyse the relationship between different software popularity measures. As a case study, for a large dataset of 175k \npm packages, we computed and extracted 9 different popularity metrics from three open source tracking systems: \librariesio, \npmjs and \github. We found that indeed popularity can be measured with different unrelated metrics, each metric can be defined within a specific context. This indicates a need for a generic framework that would use a portfolio of popularity metrics drawing from different concepts.
\end{abstract}

\begin{IEEEkeywords}
empirical analysis, popularity, software package, \npm
\end{IEEEkeywords}

\section{Introduction}

Depending on reusable software libraries is a common practice in software development~\cite{HirschfeldLammel2005}.
It enables software developers to benefit from the functionality and maturity offered by these libraries, rather than needing to reimplement it from scratch.
During the last years, the availability of very large distributions of open source projects in many domains has greatly increased this practice.
Many programming languages feature at least one package management system to facilitate the use of such projects as external libraries~\cite{decan2017empirical2}.
Such package managers automate the distribution, installation and upgrading of thousands of different software packages~\cite{mancinelli2006managing}.

For \javascript, \npm is by far the largest package manager in terms of number of hosted packages.  
Being so large, different packages frequently provide the same or similar functionality, making it challenging for developers to find and select the most appropriate package for their needs~\cite{surveyNode}. To face this challenge, a strategy that has been reported to be followed by developers is to use the package \emph{popularity} as a main factor for its selection~\cite{bogart2016break,lee2013github}. Unfortunately, the concept of popularity can be interpreted very broadly, and can be measured in different ways.

Researchers rely on popularity metrics in order to sample their analyzed datasets or study software characteristics. Different studies rely on different popularity metrics, possibly producing different or even  contradicting outcomes. 
A frequent way to characterise popularity is in terms of \emph{social} aspects. For example, famous developers (so-called ``rock stars") have been shown to have a larger influence on where their followers contribute to than ordinary developers have on their followers~\cite{lee2013github}. Moreover, when Bogart \etal~\cite{bogart2016break} interviewed OSS developers involved about the reasons behind selecting the appropriate dependencies for their software projects, most of the responses belonged to categories related to popularity and community reputation. Borges \cite{borges2018s} found through a survey with 400 Stack Overflow users that the number of stars, forks and watchers of \github projects are all considered to be very useful popularity metrics.

Popularity can also be characterized in terms of \emph{technical} aspects, such as use (\eg number of downloads) or reuse (\eg number of depending packages). Such definitions are based on the assumption that, if a package is widely (re)used, it can be trusted to be a good package. 
 
The goal of our work is therefore to raise awareness of the relationship between, and the risk of using, different software package popularity measurements. As a case study, we focus on a large set of 175k \npm packages, and analyse 9 different popularity metrics extracted from 3 different sources (\librariesio, \npmjs and \github).
We focus on one main research question: \emph{How are metrics of package popularity related to each other?}

\section{Motivation and State of the Art}
\label{sec:related}

Popularity of open source software has been the subject of study, and used as  basis for many empirical studies. In this section, we show how broad and different is the concept of popularity among researchers.

Capra \etal\cite{capra2011firms} studied the impact of company participation on popularity and software design quality for 643 SourceForge projects. Using popularity as the ranking index in SourceForge, they found that company involvement improves the popularity of open source projects.
Borges \etal\cite{borges2016understanding} measured popularity as the number of stars of \github projects. They analyzed the factors that impact this popularity, and also studied the impact of new features on project popularity. Analyzing 2,279 popular \github repositories, they identified four main patterns of popularity growth.
Sajnani \etal\cite{sajnani2014popularity} studied the relation between quality and library popularity in \maven, and considered popularity as the external usage across a set of open source \java projects. 

Inspired by the work of Dabbish \etal\cite{dabbish2012social}, Karan \etal\cite{aggarwal2014co} investigated the relation between project popularity in \github and its consistent documentation updates. They considered popularity in terms of community interest, computed as the sum of the number of stars, number of forks and number of pull requests. Using this metric they found strong indicators that consistently popular projects exhibited consistent documentation effort.
Syed \etal\cite{syed2016data} analyzed the relation between the project popularity in terms of forks, watchers, stars, pull requests and code change frequency in \github projects. They showed that projects with at least 1,500 watchers each month have a strong positive correlation between project popularity and frequency of code changes.
Kula \etal\cite{kulageneralized} considered library popularity as the number of internal dependents (\ie internal usage: how many other libraries are using it), and proposed a model to visualize library popularity.

Dey \etal~\cite{dey2018software} analyzed 13k ``popular" packages in \npm, using linear regression and random forest models to inspect the effects of predictors representing different aspects of the software dependency supply chain on changes in numbers of downloads for a package. Considering popularity as downloads, they found that the number of downloads of upstream and downstream runtime dependencies have a strong effect on the number of package downloads. This suggests that, in order to interpret the package downloads properly, one should take into account the peculiarities of both upstream and downstream dependencies of that package.

The findings of the above studies are difficult to compare because they use different metrics to evaluate popularity.
To our knowledge, there is no research focusing on how popularity measurements coming from different sources are related.
The vision of this work is to raise awareness about the usage of, and relation between, different measurements of popularity.  By doing so, we hope to gain more insights about how to understand and interpret the results of similar studies using popularity in different contexts.

\section{Method}
\label{sec:methodology}

The first step for studying popularity of software packages is to choose a relevant package management system.
We require a widely used and well-known packaging system that involves a large and active developer community, so that measures about package popularity are relevant. We also need a large number of software packages, to minimize possible bias. These requirements led us to select \npm, the largest packaging system both in number of packages (over $800k$ packages as of October 2018). 
To calculate popularity metrics for \npm packages, we extracted and combined information from three online data sources and tracking services:
\begin{enumerate}

\item \librariesio, an open source repository containing metadata of package dependencies extracted from 36 package managers.\footnote{\url{https://zenodo.org/record/1196312} (CC Share-Alike 4.0 license)}
The extracted metrics are based on the dataset of 13\textsuperscript{th} March 2018, containing $698k$ packages.

\item \npmjs, the official website for the \npm package manager, allows to search for \npm packages and sort them based on their \emph{popularity}, \emph{quality} or \emph{maintenance} characteristics.\footnote{See \url{https://docs.npmjs.com/getting-started/searching-for-packages}.}
Popularity is computed as a weighted sum of number of stars, number of forks, number of subscribers, number of contributors,
number of downloads, downloads acceleration and number of dependents.

\item \github, the hosting platform of many \npm packager repositories. 73\% of all \npm packages are hosted on \github, 0.83\% on Bitbucket, 0.69\% on Gitlab, and 24.17\% using an undefined repository. Thus, we decided to focus only on \npm packages hosted on \github.
\end{enumerate}

\begin{table*}[!ht]
	\centering
	\caption{Descriptive summary of \npm package popularity metrics.}
	\label{metrics}
\begin{tabular}{l|l|l}

\textbf{Metric }                                                                                                   & \textbf{Source }                    & \textbf{Description}                                                                                                                                                                          \\ \hline

\multicolumn{1}{l|}{\begin{tabular}[c]{@{}l@{}}\# runtime dependent repositories\end{tabular}}        & \librariesio                & \multicolumn{1}{l}{number of non-\npm repositories that have a direct runtime dependency on the package} \\ 

\multicolumn{1}{l|}{\begin{tabular}[c]{@{}l@{}}\# transitive runtime dependents \end{tabular}} & \librariesio                & \multicolumn{1}{l}{number of \npm packages that have a transitive runtime dependency on the package}                      \\ 

\multicolumn{1}{l|}{\# (direct) runtime dependents}                                                       & \begin{tabular}[c]{@{}l@{}}\librariesio \\ \npm  \end{tabular}                  & \multicolumn{1}{l}{number of \npm packages that have a direct runtime dependency on the package}                                                                     \\ 
\hline
\multicolumn{1}{l|}{\# downloads}                                                                        & \npm                        & \multicolumn{1}{l}{number of downloads of the package during the last year}                                                                                                         \\ 

\multicolumn{1}{l|}{\# npm stars}                                                                        & \npm                        & \multicolumn{1}{l}{number of stars given by npm developers (\url{https://docs.npmjs.com/cli/star})}                                                             \\ 
\hline

\multicolumn{1}{l|}{\# github stars}                                                                     & \github                     & \multicolumn{1}{l}{number of stars of the package repository}                                                                                                                       \\ 

\multicolumn{1}{l|}{\# forks}                                                                            & \github                     & \multicolumn{1}{l}{number of forks of the package repository}                                                                                                                       \\ 

\multicolumn{1}{l|}{\# pull requests}                                                                    & \github                     & \multicolumn{1}{l}{number of pull requests and issues of the package repository}                                                                                                    \\ 

\multicolumn{1}{l|}{\# subscribers}                                                                      & \github                     & \multicolumn{1}{l}{number of subscribers/watchers of the package repository}                                                                                                        

\end{tabular}
\end{table*}

Starting from the \librariesio dataset, we restrict our analysis to packages that are at least two years old, because we consider that those packages had enough time to accumulate popularity. Then, using package names we extracted package information from \npmjs. When different packages have the same \github repository (\eg \textit{lodash} and \textit{lodash-es}), we choose the popularity statistics of the package with the highest number of downloads. This occurs in 6,898 distinct packages that share 2,336 GitHub repositories.
This led us to a final set of 175,774 \npm packages for our empirical analysis, whose data was extracted by April 2018.

\section{How Are Metrics of Package Popularity Related to Each Other?}
\label{sec:evaluation}

\subsection{Dataset Exploration}

\tab{metrics} summarises the popularity metrics extracted from the different data sources used in our analysis.
For the dependency-based metrics, we only considered \textit{runtime} dependencies, \ie those required to install and execute the package. 
Indeed, when calculating the number of (direct) runtime dependents for packages from \librariesio, we obtained almost identical values as the number of dependents reported by \npmjs (with a linear correlation of $R=0.99$). The support team of \npm indeed confirmed that the metric used by \npmjs is based only on direct runtime dependencies, \ie packages that are used as \textit{development} or \textit{other dependencies} are not taken into account\footnote{For more details on dependency types: \url{https://docs.npmjs.com/files/package.json}}. 

Exploring the dataset, we found that 27\% of the \npm packages have no direct runtime dependents, 
and 62\% are not used in any external repository. We also found that only 39\% of all packages have been downloaded more than 1,000 times in the last year and 76\% have no \npm stars.
For the metrics from \github, we found that 32\% of \npm package repositories have no stars, 50\% have no forks, 54\% have only one subscriber and 43\% of the repositories have no pull requests. This suggests that ``popularity" can be quantified in many ways, obtaining different results.

To study the relation between different metrics, we perform a pairwise Pearson ($R$) and Spearman ($\rho$) correlation for all pairs of metrics. We use the following thresholds to interpret the absolute values of the correlation coefficients: 
 $0<\textit{very\: weak}\leq0.2<\textit{weak}\leq0.4<\textit{moderate}\leq0.6<\textit{moderately strong}\leq0.8<\textit{strong}\leq1$.

\subsection{Metrics Emanating From the Same Source}

To study how different metrics of popularity are related we consider the sources from where they were extracted as categories, and we study the correlation between different metrics from the same category.

\subsubsection{libraries.io}
Dependency-based metrics can be used to determine if a package is a top-level application dependency (\ie used often by external repositories), or a popular internal library (\ie used  often by other \npm packages). 
\tab{librariesio} reports the correlation results for the considered pairs of metrics from \librariesio. The correlation results vary from \textit{moderate} to \textit{moderately strong}. 
This means that choosing one of these metrics as the popularity measure will provide similar, but slightly different results than the other metrics.

\begin{table}[!ht]
\centering
\caption{Pearson and Spearman correlation coefficients for \librariesio popularity metrics}
\label{librariesio}
\begin{tabular}{l|l|l}
                                                                            & \begin{tabular}[c]{@{}l@{}}\# transitive runtime\\  dependents\end{tabular} & \begin{tabular}[c]{@{}l@{}}\# (direct) runtime \\ dependents\end{tabular} \\ \hline

\begin{tabular}[c]{@{}l@{}}\# dependent runtime\\ repositories\end{tabular} & \begin{tabular}[c]{@{}l@{}}$\rho = 0.63$\\ \textbf{\textit{moderately strong}}\\ $R= 0.58$\\ \textit{moderate}\end{tabular}                & \begin{tabular}[c]{@{}l@{}}$\rho = 0.53$\\ \textit{moderate} \\ $R= 0.71$ \\ \textbf{\textit{moderately strong}}\end{tabular}         \\ \hline   
  
  \begin{tabular}[c]{@{}l@{}}\# direct runtime \\ dependents\end{tabular}     & \begin{tabular}[c]{@{}l@{}}$\rho = 0.66$\\ $R= 0.62$ \\ \textbf{\textit{moderately strong}}\end{tabular}                &                                                                         
\end{tabular}
\end{table}

\subsubsection{npm} 

Between the metrics extracted from \npm there was little correlation (see \tab{corr_npm}).
We found a \textit{moderately strong} Pearson correlation 
between \#~npm stars and \#~runtime dependents but the Spearman correlation was weak.
The other correlations were weak to moderate.

\begin{table}[!ht]
\centering
\caption{Pearson and Spearman correlation coefficients for \npm popularity metrics. }
\label{corr_npm}
\begin{tabular}{l|l|l}
                                                                         & \# downloads                                                       & \# (direct) runtime dependents                                       \\ \hline
\# npm stars                                                             & \begin{tabular}[c]{@{}l@{}}$\rho = 0.39$\\  $R = 0.33$ \\ \textit{weak}\end{tabular} & \begin{tabular}[c]{@{}l@{}}$\rho = 0.27$ \\\textit{weak}\\ $R = 0.75$ \\\textbf{\textit{moderately strong}}\end{tabular} \\ \hline
\begin{tabular}[c]{@{}l@{}}\# (direct) runtime                                       \\  dependents\end{tabular} & \begin{tabular}[c]{@{}l@{}}$\rho = 0.42	$\\ $R = 0.45$ \\ \textit{moderate}\end{tabular} &                                                                   

\end{tabular}
\end{table}

\subsubsection{GitHub} 
Metrics from \github are important indicators of how the community of developers accepts and judges a software project. Similar to Borges \etal \cite{borges2016understanding} analysis of the factors that impact the \#~github stars, \tab{corr_github} shows a \emph{strong} correlation with the \#~forks of the package repositories.
For all pairs of \github popularity metrics, we found a moderate to moderately strong correlation, except for the pair of \textit{(\#subscribers, \#pull requests)}, where we only found moderate Spearman and Pearson correlations.

\begin{table}[!ht]
\centering
\caption{Pearson and Spearman correlation coefficients for \github popularity metrics.}
\label{corr_github}
\begin{tabular}{p{1.4cm}|p{1.75cm}|p{1.75cm}|p{1.75cm}}
                 & \#github stars                                                    & \#forks                                                             & \#subscribers                                                     \\ \hline
                 
\#pull requests & $\rho = 0.64$ \quad $R = 0.64$  \textit{\textbf{moderately strong}} & $\rho = 0.7$ \quad $R = 0.63$ \textit{\textbf{moderately strong}}  & $\rho = 0.53$ \quad $R = 0.51$ \textit{moderate} \\ \hline
                 
\#subscribers   & $\rho = 0.55$  \quad \textit{moderate}  \quad $R = 0.7$  \quad \textit{\textbf{moderately strong}} & $\rho = 0.55$ \textit{moderate}  \quad $R = 0.68$ \textit{\textbf{moderately strong}} &                                                                    \\ \hline

\#forks         & $\rho = 0.73$  \quad \textit{\textbf{moderately strong}} \quad $R = 0.85$ \quad \textit{\textbf{strong}} &                                                                      &                                                                    

\end{tabular}
\end{table}

\subsection{Metrics Emanating from Different Sources}

To limit the number of popularity metrics to consider, we focus on 6 metrics: \textit{\#dependent repositories} from \librariesio, all three metrics from \npmjs, \textit{\#subscribers} and a single popularity metric --defined and used by Aggarwal \etal~\cite{aggarwal2014co}-- that combines the three \github metrics:
$$\textit{Aggarwal-Popularity} = \#\textit{forks} + \#\textit{stars} + \#\textit{pull requests}^2$$

\begin{figure}[!ht]
	\begin{center}
		\setlength{\unitlength}{1pt}
		\footnotesize
		\includegraphics[width=0.95\columnwidth]{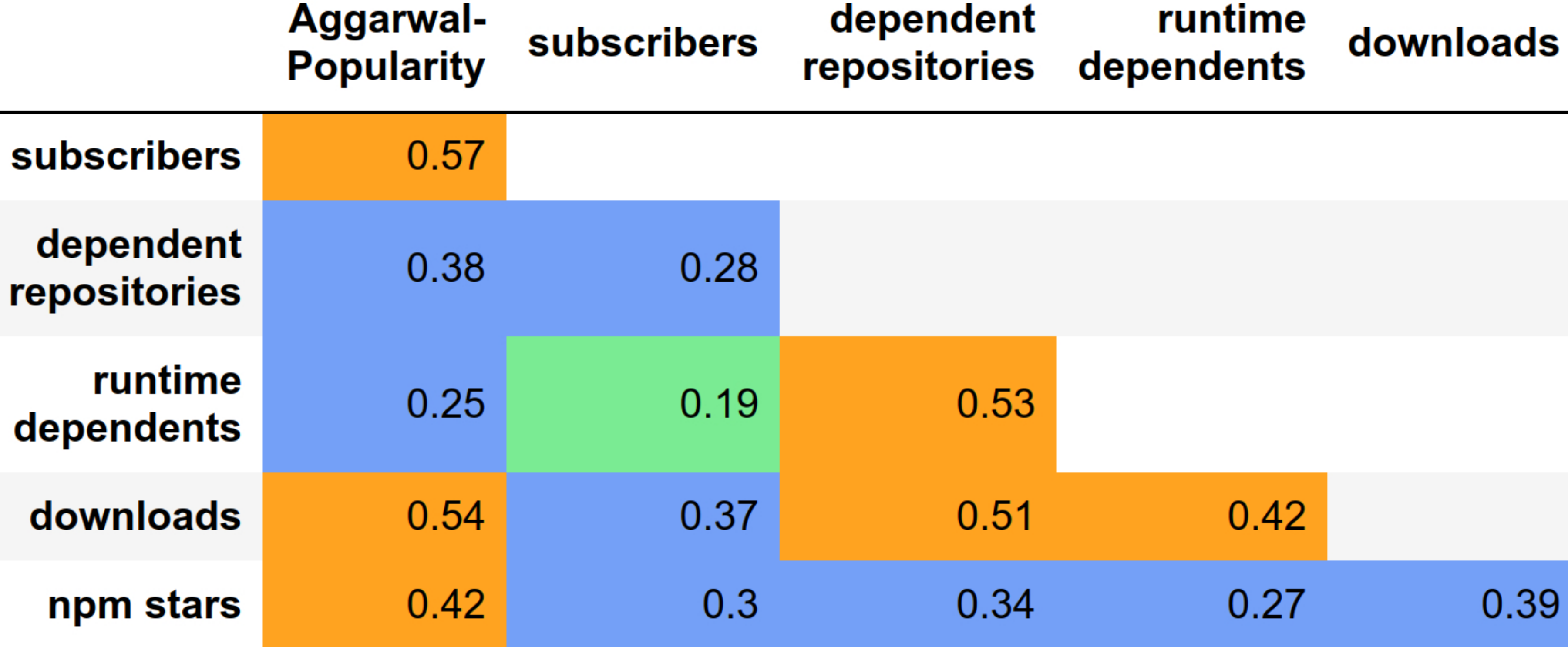}
		\caption{Spearman correlation between popularity metrics. (Orange background = moderate correlation; blue = weak; green = very weak)}
		\label{fig:correlation}
	\end{center}
\end{figure}

\fig{fig:correlation} reports the Spearman correlation between these 6 metrics, revealing that the highest correlation is only \emph{moderate} ($\rho=0.54$), for the pair (\textit{Aggarwal-Popularity}, \textit{\#downloads}).

To study the relation between the 6 metrics in more detail we use the \textit{\# direct runtime dependents} metric as a main sorting criterion to select a set of 1,000 packages out of our larger dataset, thereby simulating a search on the  \npm home page that by default retrieves the most depended upon packages\footnote{https://www.npmjs.com/browse/depended}.
For the other 5 popularity metrics we compute the sets of top 1,000 packages in a similar way, and we compare all possible intersections between these sets and the one based on the main sorting criterion.

\begin{figure}[!ht]
	\begin{center}
		\setlength{\unitlength}{1pt}
		\footnotesize
		\includegraphics[width=0.95\columnwidth]{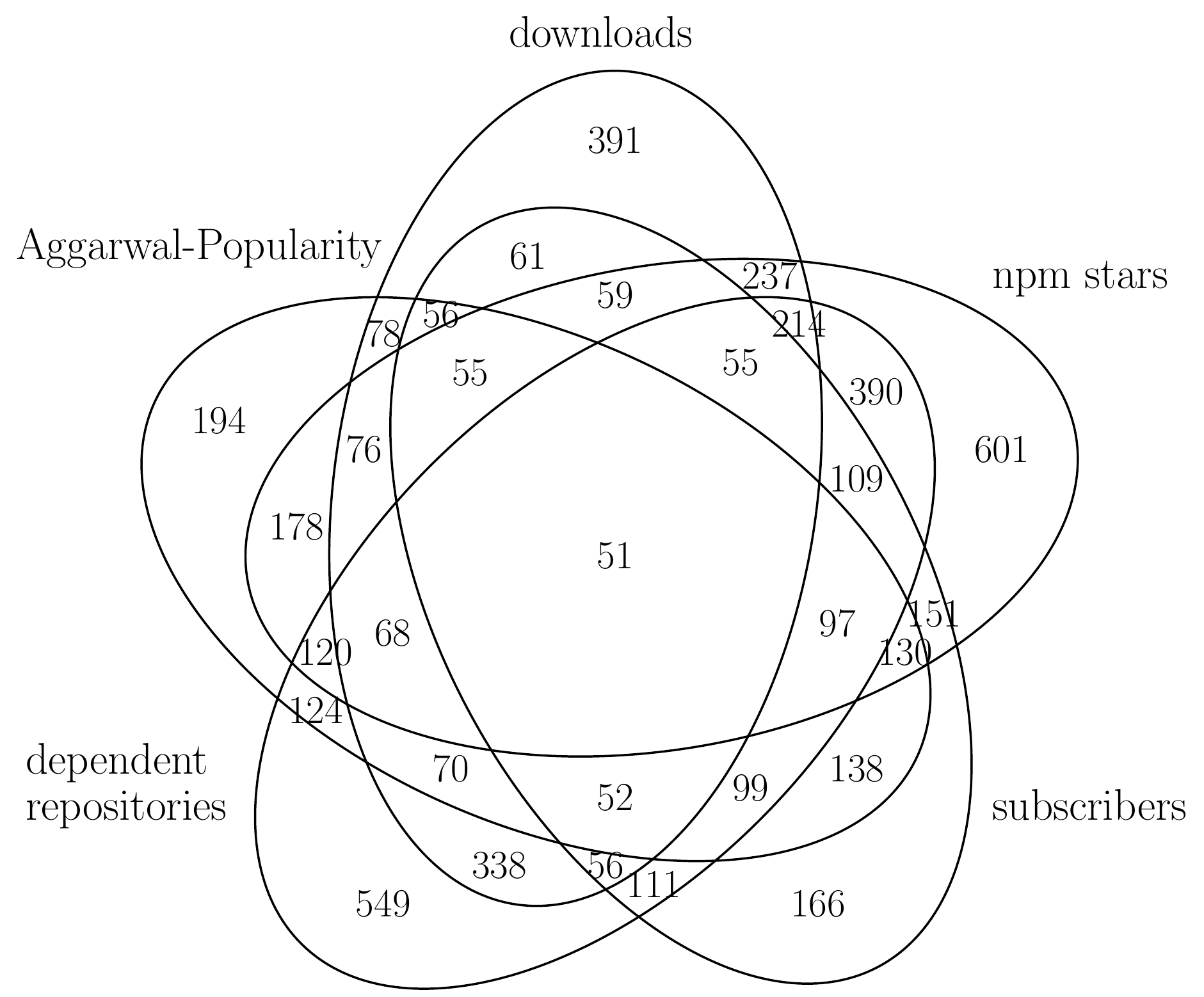}
		\caption{Number of packages from the top 1,000 most depended upon packages, that are in the top 1,000 in terms of other popularity metrics. }
		\label{fig:venn_pop}
	\end{center}
\end{figure}

The results are shown using a \textit{Venn diagram} in \fig{fig:venn_pop}. 
If we consider the intersection of \emph{all} popularity metrics, we find only 5.1\% (\ie 51 out of 1,000) packages.
If we compare the \textit{\# direct runtime dependents} metric with each of the 5 reported metrics individually, we obtain the largest intersections with \textit{\#npm stars} (601 packages), and \textit{\#dependent repositories} (549 packages).
The smallest intersection is found for \emph{\#subscribers} (166 packages) and \emph{Aggarwal-Popularity} (194 packages), that have less than 20\% in common with the \textit{\# direct runtime dependents} selection.
The reported numbers are in general low, indicating that the main sorting criterion generates many different packages than the other popularity metrics.

\section{Discussion}
\label{sec:discussion}

Our empirical analysis revealed that most of the considered popularity metrics are not strongly correlated to each other. Even if they are, as was the case for \#~forks versus \#~github stars (see \tab{corr_github}), other authors have found lower correlations. For example, \cite{borges2018s} only found a moderate correlation ($\rho=0.558$), showing the influence of the chosen population\footnote{They relied on GitHub's 5000 most starred repositories.} on the outcome of the results.

Popularity is indeed relative and should be always measured to a specific context. This finding should be taken into account when designing automated recommendation tools and search engines for software libraries. For example, the \npmsio search engine for \npm packages, can sort results based on popularity, quality, maintenance, or a combination of those. The popularity search criterion is based on a weighted sum of 7 more primitive metrics. Using other weights or alternative definitions of popularity may produce quite different popularity rankings.

Implementations of popularity metrics might also produce incorrect values, \eg because they were computed incorrectly, or because the data has been ``violated" by spammers. For example, in 2016 an application was created\footnote{https://github.com/ell/npm-gen-all} to generate many spam \npm packages (\eg neat-x and wowdude-x\footnote{https://libraries.io/npm/wowdude-119}) that used a huge number of dependencies. This affected the \#~downloads and \#~dependent packages for those packages that were used as dependencies. 
Software maintainers are raising awareness about such \emph{download inflation} and related issues, \eg \textit{``What happens if a large team is using the library and now has to download it for every single one of their users? More download inflation."\cite{tideliftDownloads}}.

The above phenomena may impact reproducibility of research studies if the population of software libraries under analysis is retrieved using a single metric of popularity (\eg \cite{borges2018s} analysed the top 5,000 \github projects based on their number of stars). Other popularity metrics may lead to different research findings, limiting the generalisability of the results.

All of the above issues are important to consider if the goal is to use or analyse popularity metrics across software libraries with different characteristics. Thus, there is an urgent need for a popularity measurement framework  including a wide range of popularity metrics, as previous studies have done with other open source software characteristics (\eg software reusability \cite{ampatzoglou2018reusability}, software quality \cite{Robinson2011} and software success \cite{crowston2003defining}). 

Our findings are valuable for the open source community at large. For instance, the Linux Foundation's  \textit{CHAOSS} project focuses on creating analytics and metrics to help define community health. One of its goals is to define metrics for measuring community activity, contributions, and health, and eventually produce detailed use cases or recommendations to analyze specific issues in the industry/OSS world\footnote{https://chaoss.community/metrics/}. Our analysis could be a contribution to the project in order to define the most important metrics for popularity.

\section{Threats to Validity}
\label{sec:threats}

The reported correlation results may be biased by the selected dataset, and the filtering to only \npm packages hosted on \github with at least two years of history. To limit this bias we analyzed more than 175k \npm packages. 
Our results may not be generalizable to \javascript packages that are not distributed through \npm, that are not hosted on \github, or that are published by other software package managers. We cannot generalise either to packages that are developed in other programming languages.

Another threat to generalisability is that we have restricted the dependency analysis to runtime dependencies only. Hence, the popularity of packages that are frequently or exclusively used as development dependencies has been underestimated.

Many other popularity metrics could have been considered in the study (\eg \librariesio SourceRank\footnote{https://github.com/librariesio/libraries.io/issues/1916}).
Not including them in the analysis may have influenced the results of our analysis.

\section{Conclusion and Future Work}
\label{sec:conclusion}

Software popularity is an important indicator of software success. It can be a main factor for gaining more attention and adoption, and it may attract new developers to the project. However, which characteristic of software popularity to be considered should be defined by the specific context of use.

In this paper, we showed that popularity can be measured in many different ways, and researchers have studied popularity using a wide variety of different metrics. Their research findings may depend on the metrics they used. To illustrate this, we relied on a large dataset of \javascript packages in \npm to study the correlation between different popularity metrics used by different open source services. We observed that many popularity metrics are not strongly correlated, implying that the use of different metrics may produce different outcomes. This calls for the definition and use of a measurement framework that takes into account the diversity and context-dependence of software popularity.

As future work, we plan to complement our quantitative analysis with qualitative interviews of how developers actually use popularity measures, and whether they are concerned about the issues that relying on such measures may bring. The outcome of such an analysis may ultimately lead to better services for searching and recommending software libraries.

We also plan to reproduce existing research studies in order to assess to which extent the use of specific popularity metrics affects the research findings.
In a similar vein, we aim to extend our analysis to package managers for other programming languages 
(\eg PyPI for Python, Maven for Java, RubyGems for Ruby).

\section*{Acknowledgment}
This work was partially supported by the EU Research FP (H2020-MSCA-ITN-2014-642954, \textsf{Seneca}), the Spanish Government (TIN2014-59400-R, \textsf{SobreVision}), the Excellence of Science Project \textsf{SECO-Assist} (O015718F, FWO - Vlaanderen and F.R.S.-FNRS).

\providecommand{\noopsort}[1]{}

\end{document}